\documentclass[sigconf]{acmart}
\AtBeginDocument{%
  }

\usepackage{fancyhdr}
\usepackage{multirow}
\usepackage{listings}
\copyrightyear{2026}
\acmYear{2026}
\setcopyright{cc}
\setcctype{by}

\acmConference[L@S '26]{Proceedings of the Thirteenth ACM Conference on Learning @ Scale}{June 29-July 03, 2026}{Seoul, Republic of Korea}
\acmBooktitle{Proceedings of the Thirteenth ACM Conference on Learning @ Scale (L@S '26), June 29-July 03, 2026, Seoul, Republic of Korea}
\acmDOI{10.1145/3774398.3811616}
\acmISBN{}

\begin{document}


\title[Teachers' Perceived Benefits and Risks of AI Across Fifty-Five Countries: \\ An Audit of LLM Alignment and Steerability]{Teachers' Perceived Benefits and Risks of AI Across Fifty-Five Countries: An Audit of LLM Alignment and Steerability}

\titlenote{\scriptsize
\textcopyright\ 2026 Copyright held by the owner/author(s). Publication rights licensed to ACM.
This is the authors' preprint version of the work. It is posted here for personal use only.
Not for redistribution. The definitive Version of Record will appear in the
Proceedings of the Thirteenth ACM Conference on Learning @ Scale (L@S '26),
July 27--31, 2026, Seoul, Republic of Korea.
}
\author{Yan Tao}
\affiliation{%
  \institution{Cornell University}
  \city{Ithaca}
  \state{NY}
  \country{United States}
}
\email{yt589@cornell.edu}

\author{Olga Viberg}
\affiliation{%
  \institution{KTH Royal Institute of Technology}
  \city{Stockholm}
  \country{Sweden}
}
\email{oviberg@kth.se}

\author{Deepak Varuvel Dennison}
\affiliation{%
  \institution{Cornell University}
  \city{Ithaca}
  \state{NY}
  \country{United States}
}
\email{dv292@cornell.edu}

\author{Zhikun Wu}
\affiliation{%
  \institution{Linköping University}
  \city{Linköping}
  \country{Sweden}
}
\email{zhikun.wu@liu.se}

\author{René F. Kizilcec}
\affiliation{%
  \institution{Cornell University}
  \city{Ithaca}
  \state{NY}
  \country{United States}
}
\email{kizilcec@cornell.edu}

\renewcommand{\shortauthors}{Yan Tao, Olga Viberg, Deepak Varuvel Dennison, Zhikun Wu, René F. Kizilcec}

\begin{abstract}
Teachers’ trust in artificial intelligence (AI) in education depends on how they balance its perceived benefits and risks. Yet global discussions about scaling AI in education rely on fragmented evidence, as most studies of teachers’ perceptions focus on single countries or small samples. This lack of representative cross-national evidence limits both theory building and policy development. At the same time, large language models (LLMs) are increasingly used in research, policy, and teachers’ professional workflows, despite limited validation in education.
To address these gaps, we conduct a large-scale audit of LLM alignment with teachers’ perceptions of AI by combining representative international survey data with systematic model evaluation. Using OECD TALIS data from 55 countries and territories, we measure cross-national variation in teachers’ perceived benefits and risks of AI. We then benchmark responses from eight state-of-the-art LLMs across four providers under both general and country-specific prompting, comparing higher- and lower-reasoning models.
Results reveal substantial cross-national variation in teacher perceptions that is not reliably reflected in LLM outputs. Models compress country differences, overestimate both benefits and risks, and show limited gains from identity prompting or enhanced reasoning. This misalignment matters because LLM-generated guidance and professional discourse increasingly shape how teachers learn about and discuss AI, potentially influencing trust and future adoption decisions. Our findings caution against treating LLM outputs as substitutes for direct engagement with teachers when informing global AI-in-education initiatives. At the same time, some models (e.g., Gemini 3 Fast) partially capture cross-national ranking patterns, suggesting a complementary role in hypothesis generation and exploratory comparative analysis.
\end{abstract}

\ccsdesc[500]{Applied computing~Education}
\keywords{Teachers, Culture, Artificial Intelligence, Alignment, Human Perspectives, Education}


\maketitle

\section{Introduction}

Artificial intelligence (AI) has rapidly entered educational systems worldwide, including K-12 settings, supporting a range of teaching and learning activities such as instructional planning, automated grading, personalized learning, and administrative support \cite{huang2025artificial, tripathi2025teaching}. While advances in AI technology are critical for enabling these applications, their effectiveness in promoting teaching and learning ultimately hinges on teachers, who determine whether, when, and how AI is integrated into classroom practice \cite{kizilcec2024advance}. Teachers’ perceptions of AI, particularly how they weigh its potential benefits against its risks, constitute the central decision mechanism shaping trust, adoption, and responsible use in education \cite{viberg2024trust}.

Prior research has established that teachers’ trust in AI-based educational technologies is grounded in two core judgments: perceived benefits and perceived risks \cite{nazaretsky2022teachers}. Teachers may recognize AI’s potential to support instruction, personalization, and efficiency, while simultaneously expressing concerns about reliability, pedagogical appropriateness, bias, academic integrity, and student data privacy \cite{tan2025artificial, garcia2025perceptions, gouseti2025ethics, taheri2025factors, karran2025multi, yin2025responsible}. Importantly, these judgments are decision-relevant rather than merely attitudinal: they predict teachers’ willingness to accept vulnerability by adopting AI tools and to meaningfully integrate them into instructional practice. However, existing empirical evidence on teachers’ perceptions of AI is largely fragmented, with most studies focusing on single countries, small teacher samples, or limited sets of contexts (e.g.,~\cite{filiz2025teachers, kim2025perceptions, calleja2025primary}) .

Cross-national evidence is particularly important in education, where institutional arrangements, professional norms, and policy environments vary substantially across countries. Teachers’ perceptions of AI are shaped not only by individual characteristics but also by broader cultural and educational contexts \cite{viberg2024trust}. Without large-scale, representative international data, it remains difficult to disentangle generalizable patterns in teachers’ benefit–risk assessments from context-specific concerns tied to national education systems. This lack of comparative evidence limits both theory building and policy development, as global discussions about educational AI risk being informed by fragmented or unrepresentative findings.


At the same time, large language models (LLMs) are increasingly used to model, simulate, or extrapolate human opinions at scale (e.g., \cite{jiang2025simulating, matz2024potential, yao2024value}). In education research, policy analysis, and system design, LLMs are often treated, explicitly or implicitly, as informational substitutes for stakeholder perspectives, including those of teachers \cite{delikoura2025superficial}. This practice relies on a critical assumption of alignment: that LLM-generated responses meaningfully reflect how relevant human groups think about AI. Yet, this assumption has rarely been tested in education using empirically grounded benchmarks, particularly for domain-specific, decision-relevant judgments such as \textit{perceived benefits} and \textit{risks}. Without evidence of alignment, research and policy that rely on LLM outputs risk producing misleading or overly generalized conclusions, potentially prompting international organizations and education systems to promote narratives that overlook locally salient concerns \cite{holmes2023guidance}.

The misalignment between LLM outputs and teacher perceptions has further practical implications. From the teacher's perspective, their understanding of AI is dynamic, evolving through direct interaction with AI tools and professional experiences. When LLM-generated conversations about AI use in education (e.g. teaching coaching \cite{wang-demszky-2023-chatgpt}, lesson planning \cite{10.1145/3654777.3676390}) diverge significantly from teachers' evolving perceptions of AI's usefulness and risks in education, it may create confusion, reduce trust, and inadvertently shape their future adoption decisions. Moreover, as generative AI becomes embedded in social media and other platforms that support professional communication \cite{lu2025understanding}, misalignment may distort discussions about AI use in education, misrepresenting real-world teacher views and potentially affecting broader narratives about AI adoption in schools.

In this paper, we address the lack of cross-national evidence on teachers’ perceptions of AI for education and the alignment between LLM representations and these teacher perspectives by combining large-scale education survey analysis with audit-style evaluation methods from AI alignment research. Using data from the OECD Teaching and Learning International Survey (TALIS\footnote{OECD (2025), \textit{Results from TALIS 2024: The State of Teaching, TALIS}, OECD Publishing, Paris, https://doi.org/10.1787/90df6235-en.}
), we examine the cross-national variation in teachers’ perceived benefits and risks of AI in education across 55 countries and territories. We then benchmark responses from eight state-of-the-art LLMs from four major providers against the same survey items, comparing each provider’s latest reasoning and non-reasoning models. Models are prompted \textit{with} and \textit{without} country-specific teacher identities, enabling a systematic assessment of alignment and steerability.

Our work addresses the following four research questions:\\
\textbf{RQ1 (Cross-national variation).} To what extent do teachers’ perceived benefits and perceived risks of AI in education vary across countries?\\
\textbf{RQ2 (LLM alignment).} To what extent do LLMs’ responses align with teachers’ observed perceptions of AI benefits and risks across countries?\\
\textbf{RQ3 (Steerability through prompting).} Does prompting LLMs to respond as teachers from specific countries improve their alignment with local teacher perceptions of AI benefits and risks?\\
\textbf{RQ4 (Role of reasoning).} Do LLMs with explicit reasoning capabilities exhibit higher alignment or improved steerability compared to non-reasoning models from the same provider?

This paper makes four contributions. First, we provide a large-scale, cross-national analysis of teachers’ perceived benefits and risks of AI in education using representative TALIS data, documenting substantial variation across countries. Second, we conduct an audit-style evaluation of LLMs by benchmarking their responses to the same survey items against observed teacher distributions along decision-relevant dimensions of AI trust. Third, we systematically compare eight state-of-the-art LLMs from four major providers, enabling a robust assessment of alignment across contemporary model families. Fourth, by comparing reasoning and non-reasoning models within providers, we examine whether explicit reasoning capabilities improve alignment and steerability when models are prompted to reflect country-specific teacher perspectives.

\section{Related Work}

\subsection{Teacher Perceptions of AI: Benefits, Risks, and Trust}
Teachers’ perceptions are central to the adoption and impact of AI-enabled educational technology (AI-EdTech), because teachers ultimately decide whether and how AI is integrated into classroom practice \cite{kizilcec2024advance}. Prior work emphasizes that teachers’ willingness to accept vulnerability when adopting AI-EdTech can be conceptualized as \emph{trust}, and that trust is shaped by teachers’ assessments of \emph{anticipated benefits} versus \emph{perceived concerns/risks} \cite{viberg2024trust}. This benefit-risk structure is consistent with broader models of trust and technology adoption, and has motivated recent measurement work that operationalizes teachers’ perceived benefits and concerns as key antecedents of trust in AI-EdTech \cite{viberg2024trust}.

Empirical studies and reviews identify a range of potential teacher-facing benefits of AI-EdTech, including support for planning, implementation, and assessment \cite{celik2022promises, celik2025co}. At the same time, teachers raise concerns about reliability, appropriateness, and broader educational consequences \cite{viberg2024trust, shen2025psychological}. A recurring theme in this literature is that teachers’ AI understanding and self-efficacy are associated with stronger perceived benefits, fewer concerns, and greater trust \cite{viberg2024trust}. However, much of the evidence base remains fragmented across contexts and tool types, with limited large-scale cross-national evidence grounded in representative samples \cite{zawacki2019systematic,holmes2022state}.

\subsection{Cross-National and Cultural Variation in Perceptions of AI}

Teachers’ perceptions of AI-EdTech are expected to vary across countries, reflecting differences in educational systems, professional norms, governance, and broader societal attitudes toward technology and automation. Multi-national evidence is particularly important because technology adoption and trust in automation have long been shown to differ across cultural contexts \cite{leidner2006review,chien2018culture}. In education, recent work explicitly calls for more cross-cultural research on teachers’ trust in AI-EdTech and associated factors \cite{viberg2024trust}.

Existing cross-national studies in this area have typically involved modest samples and a limited number of countries. For example, a six-country teacher survey study found that geographic and cultural differences were associated with teachers’ trust in AI-EdTech, while also highlighting the role of AI self-efficacy and understanding \cite{viberg2024trust}. Complementing these results, research on Swedish teachers’ understanding of AI underscores how national context shapes teachers’ interpretations of AI’s educational implications \cite{velander2023swedish}. Taken together, these studies motivate the need for large-scale, representative cross-national measurement that can more reliably characterize variation in teachers’ perceived benefits and risks of AI in education.
\subsection{LLM Alignment, Cultural Bias, and Steerability}

Large language models (LLMs) are increasingly used to generate human-like responses and are sometimes treated as proxies for stakeholder perspectives in research and design. This raises a fundamental alignment question: whether LLM outputs reflect empirically observed human judgments in specific populations and domains. Recent work proposes and demonstrates \emph{disaggregated evaluation} (also described as algorithmic auditing) as a principled way to assess model behavior by benchmarking outputs against established survey measures \cite{10.1145/3531146.3533233, barocas2021disaggregated,sandvig2014auditing}. 

Large-scale audit of cultural values has compared LLM responses to nationally representative survey data and found systematic cultural bias, with model outputs resembling values common in English-speaking and Protestant European countries \cite{tao2024cultural, atari2023which, cao-etal-2023-assessing}. Notably, Tao et al. tested \emph{cultural prompting} (instructing the model to answer as a person from a target society) as a control strategy and found that, for more recent models, such prompting improved alignment for many countries \cite{tao2024cultural}. Related work in NLP also investigates cross-cultural differences and cultural bias in LLMs \cite{arora2023probing,naous2023beer}.

However, cultural alignment in abstract values does not necessarily imply alignment in domain-specific, decision-relevant judgments. Education is a particularly consequential domain because biases in model outputs can misrepresent stakeholder concerns that shape trust and adoption, and because education policy frequently depends on cross-national comparisons. This motivates auditing LLMs against representative teacher survey data on perceived benefits and risks of AI in education, and testing whether country-specific prompting can steer models toward empirically observed teacher perspectives.

Finally, recent research has highlighted the role of reasoning capabilities in LLMs for producing outputs that approximate human-like judgment and support multi-step inference \cite{khamassi2024strong, fu2023chainofthoughthubcontinuouseffort, huang-chang-2023-towards}. Reasoning models generate texts assembling intermediate steps that simulate the human thinking process, which could plausibly improve alignment with human values and perceptions by mitigating harmful cognitive shortcuts and surfacing more relevant knowledge acquired during training (and via web‑search when external tools are integrated). Empirical evidence shows that incorporating such reasoning structures can enhance LLM performance on complex problems, including logical deduction, algorithmic reasoning, causal inference, and mathematics \cite{openai2024learning, zhou2024selfdiscoverlargelanguagemodels}. However, whether reasoning consistently improves or sometimes degrades human-LLM alignment in latent professional judgments, rather than objective factual knowledge, remains an open empirical question. Our work contributes evidence on this question by comparing reasoning and non-reasoning models within providers using the same survey-based audit framework.

\section{Methods}

\subsection{Survey Data and Context}
We draw on data from the OECD Teaching and Learning International Survey (TALIS), a large-scale, internationally standardized survey focused on lower secondary education teachers and school leaders, with optional participation for primary and upper secondary education \cite{oecd2025talis, oecd2024talis_technical}. TALIS is designed to produce nationally representative estimates of teachers’ beliefs, practices, and working conditions across participating education systems. The survey employs a two-stage stratified sampling design, first sampling schools and then teachers within schools, with survey weights provided to support population-level inference.

Our analysis uses items from TALIS 2024 \cite{oecd2025talis} that assess teachers’ perceptions of AI in education. These items were administered to 278,383 in-service teachers across 55 participating countries and territories (hereafter referred to as ``countries'') during the 2023-2024 TALIS data collection cycle. All analyses use teacher-level responses and apply the recommended survey weights to ensure national representativeness. As a result, our estimates reflect nationally representative views of lower secondary school teachers in each participating country and, in countries that opted into additional coverage, also incorporate responses from primary and upper secondary school teachers, rather than convenience samples or early adopters of educational technology.

\subsection{Measures}

\paragraph{Perceived benefits of AI in education.}
Teachers’ perceived benefits of AI in education were measured using five items from ``\textit{TT4G35}'', the 35th question of the TALIS teacher questionnaire. These items assess the extent to which teachers agree that artificial intelligence:
(A) helps teachers write or improve lesson plans; 
(B) enables teachers to adapt learning material to different students’ abilities; 
(C) assists teachers in supporting students individually; 
(D) supports students with specific needs (e.g., multilingual learners, students with special education needs); and 
(E) helps teachers automate administrative tasks.
Responses were recorded on a four-point Likert scale ranging from \emph{strongly disagree} to \emph{strongly agree}, with an additional \emph{I don’t know} option treated as missing. 

We conceptualize these items as reflecting a common latent construct capturing teachers’ perceived instructional and professional benefits of AI. The items demonstrated high internal consistency (Cronbach’s $\alpha = \text{0.90}$ overall), and reliability estimates were stable across countries ($\alpha$ ranged from 0.78 to 0.94). To construct a benefits index, we averaged responses across the five items, with higher values indicating greater perceived benefits of AI in education.

\paragraph{Perceived risks of AI in education}
Teachers’ perceived risks of AI in education were measured using the remaining five items from the same TALIS question. These items capture concerns that artificial intelligence:
(F) enables students to misrepresent others’ work as their own;
(J) makes recommendations that may not be appropriate or correct;
(H) amplifies biases that reinforce students’ misconceptions;
(I) jeopardises the privacy and security of student data; and
(J) suggests unsuitable pedagogical approaches that teachers would use with students.
The response scale and treatment of missing responses were identical to those used for the benefits items.

These items were conceptualized as representing a coherent construct reflecting pedagogical, ethical, and epistemic risks associated with AI use in education. The scale exhibited high internal consistency (Cronbach’s $\alpha = \text{0.85}$ overall), with consistent reliability across countries ($\alpha$ ranged from 0.72 to 0.88). We constructed a risks index by averaging responses across the five items, with higher values indicating greater perceived risks of AI in education.

\subsection{LLM Response Generation Pipeline}

To evaluate the alignment between teachers’ survey responses and large language model outputs, we generated LLM-based responses to the same TALIS survey items. We selected eight state-of-the-art LLMs from four major foundation model providers as listed in Table \ref{table:models}.  For each provider, we selected the most recent model and tested both high-reasoning and low/no-reasoning (hereafter ``low'') configurations, holding the model constant to isolate reasoning effects. For xAI, two separate Grok 4 Fast variants were used for reasoning and non-reasoning conditions, as configuring reasoning within a single variant was not possible at the time of experimentation.
All models were queried using default API settings ($T = 1.0$; provider-defined thinking budgets where applicable) to capture out-of-the-box behavior and minimize confounds. Reasoning traces were recorded where accessible (Google Gemini and Anthropic Claude), but were unavailable for OpenAI GPT and xAI Grok.
\begin{table}[h]
\centering
\caption{Model \& Reasoning Capabilities by Provider}
\label{table:models}
\begin{tabular}{lll}
\toprule
\textbf{Provider} & \textbf{Model} & \textbf{Reasoning} \\
\midrule
\multirow{2}{*}{Google}
 & Gemini 3 Flash & High \\
 & Gemini 3 Flash & None \\
\hline
\multirow{2}{*}{OpenAI}
 & GPT 5.2 & High \\
 & GPT 5.2 & None \\
\hline
\multirow{2}{*}{Anthropic}
 & Claude 4.5 Haiku & High \\
 & Claude 4.5 Haiku & Low \\
\hline
\multirow{2}{*}{xAI}
 & Grok 4 Fast Reasoning & High \\
 & Grok 4 Fast Non-reasoning & None \\
\bottomrule
\end{tabular}
\end{table}

\paragraph{Prompting procedure.}
Each model was prompted to respond to the TALIS AI-related survey items using a standardized prompt that mirrored the original survey wording. In the baseline condition, models were instructed to answer each item as a teacher responding to a survey about artificial intelligence in education (``\textit{You are a primary or secondary school teacher responding to a survey about the use of artificial intelligence (AI) in education. Please indicate your level of agreement with the list of statements presented in the survey...}''). The full prompt can be found in our OSF project: \url{https://osf.io/6mu92}. This prompt design followed established practice in LLM alignment auditing, where the model is instructed to answer survey questions from the perspective of a specified human population~\cite{tao2024cultural, cao-etal-2023-assessing, Santurkar_23_whoseopion}. The teacher identity was framed as a primary or secondary teacher role to ensure cross-national interpretability, avoiding grade-level distinctions that may be inconsistently defined across education systems and ambiguously interpreted by LLMs. 
As a robustness check, we verified that baseline responses are consistent with a more narrow prompt formulation identifying teachers as \textit{secondary school teachers teaching grades 6 to 9}'', which characterizes the majority of survey respondents. The absolute differences in perceived benefit and risk scores between the two versions were small for nearly all model–reasoning combinations: all were below 0.1 except for Grok 4 Fast in high reasoning (0.2), and 11 of the 16 comparisons were smaller than 0.05. 



In the country-specific prompting condition, the prompt additionally specified a national teacher identity (e.g., \textit{“You are a primary or secondary school teacher from \textbf{Sweden} responding to a survey about the use of artificial intelligence (AI) in education.”}). Aside from the country identifier, all prompts were held constant across models and conditions to ensure comparability. This condition drew on prior research~~\cite{tao2024cultural} demonstrating that inserting a country or nationality identity can help increase LLM alignment with human cultural values, aiming to test whether and how this lightweight strategy generalizes to teacher-LLM alignment in AI perceptions.

We evaluated two response generation approaches: generating responses for each question statement using separate prompts and generating responses for all statements within a single prompt. As no meaningful differences were observed between the two approaches, we used the single-prompt approach for the final runs to improve token efficiency. Responses were validated using a JSON schema to ensure adherence to the required survey format. The schema required the model to output the statement identifier, the full statement text, and the corresponding response for all items, ensuring alignment between statements and responses and reducing hallucinated or misaligned outputs. 


\paragraph{Replication and aggregation.}
To assess response stability, we generated ten independent responses per model, reasoning capability, and prompt condition (baseline and country-specific). We pre-committed to expanding replications if substantial variability emerged, but observed minimal variation across runs. In the baseline condition, 6 of the 8 model-reasoning combinations showed zero variance across ten runs for at least 7 of the 10 perception items. 
Under country-specific prompting, over half of model-reasoning-country combinations showed zero variance for at least 7 of the 10 items, and 93\% had at most 2 differing responses for at least 8 of the 10 items. This limited variance indicated that additional runs would yield highly similar results. We therefore calculated the perceived benefit and risk indices for each run, then averaged these scores across the ten runs for each model-reasoning-country combination (or model-reasoning combination) to represent LLM perceptions of AI for each country (or the baseline condition).
\subsection{Analytical Approach}

Our analysis proceeded in four steps. First, we examined cross-national variation in teachers’ perceived benefits and risks of AI using the TALIS survey data. ``\textit{I don't know}'' responses were recoded as missing values and analyzed separately. For each teacher, we calculated their perceived benefit scores by averaging responses to items \textit{TT4G3A}-\textit{E} and perceived risk scores by averaging responses to items \textit{TT4G35F}-\textit{J}, excluding missing values. These individual-level scores were then aggregated to weighted country-level means, excluding individuals with all missing values for a given dimension. We followed TALIS guidelines~\cite{oecd2024talis_technical} and used teacher sampling weights (\textit{TCHWGT}) to calculate weighted means and variances, ensuring nationally representative estimates rather than sample means.
Second, we assessed LLM-human alignment by comparing LLM-generated responses in the baseline condition with teacher distributions at both the country level and across the benefit and risk dimensions. 
Third, we evaluated steerability of specifying a national teacher identity in the LLM prompt by comparing the distribution of LLM-human differences under baseline versus country-specific prompting, with paired Wilcoxon signed-rank tests to statistically assess shifts in LLM-human alignment. We also conducted Spearman correlation analyses on LLM representations of each country's teacher perceptions and the corresponding country-level estimate from the teacher survey to assess whether LLMs under country-specific prompting could capture the relative ranking of countries in teachers' perceived benefits and risks. 
Forth, we examined whether high-reasoning LLMs exhibited systematically different alignment patterns than low-reasoning models. We compared the distribution of LLM-human differences for high versus low reasoning models from the same provider and conducted both paired t-tests and Wilcoxon signed-rank tests on the absolute differences between country-prompted LLM responses and teacher survey responses.

\section{Results}
\subsection{Cross-National Variation in Teachers’ Perceived Benefits and Risks of AI}

\begin{figure*}[t]
    \centering
    \includegraphics[width = 0.9\textwidth]{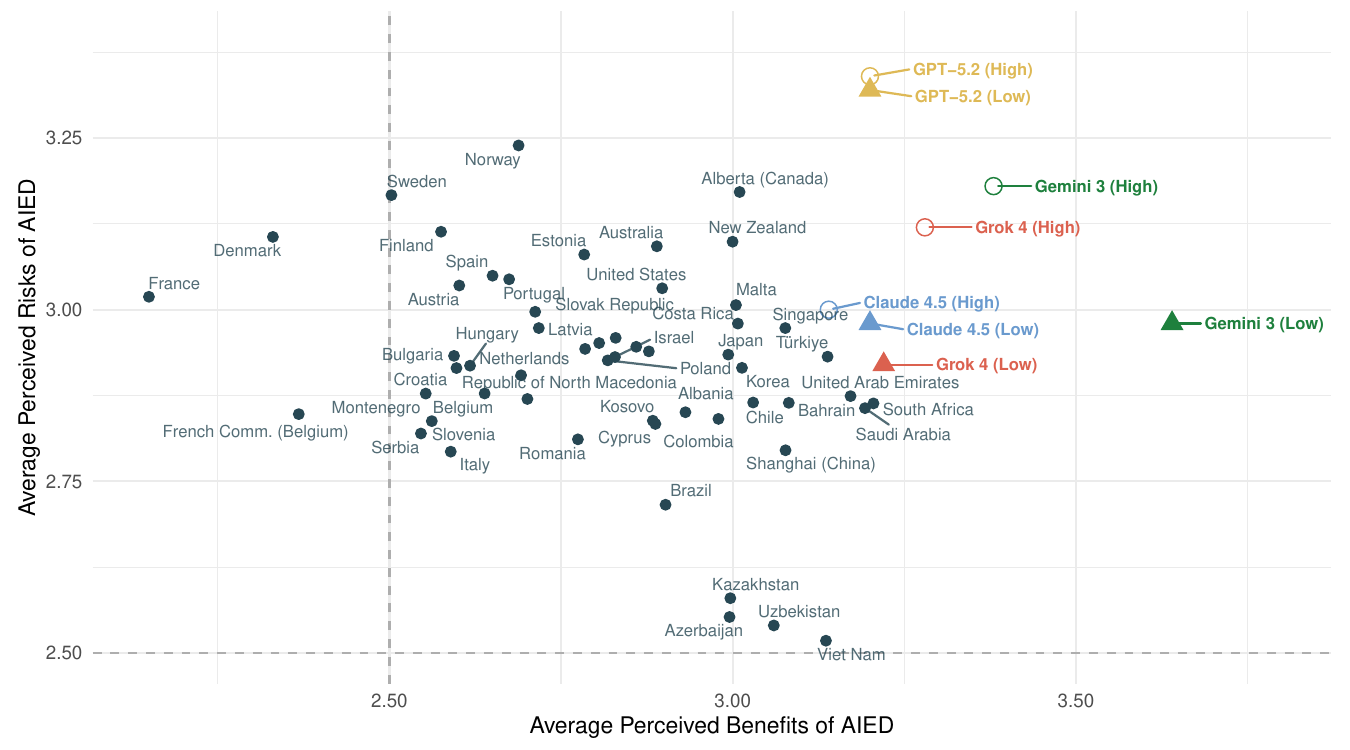}
    \caption{Teacher benefit and risk perceptions of AI for education across 55 countries/territories. Black points represent country-level mean benefit (x-axis) and risk (y-axis) ratings on 4-point scales (1 = Strongly Disagree to 4 = Strongly Agree). Dashed lines at x = 2.5 and y = 2.5 indicate the theoretical scale midpoints. Colored shapes represent averaged baseline LLM responses from four providers: triangles denote low-reasoning models and hollow circles denote high-reasoning models.}
    \label{fig:2dimension_plot}
\end{figure*}

Teachers’ perceptions of the benefits and risks of AI in education varied substantially across countries, as shown in Figure~\ref{fig:2dimension_plot}. Both the perceived benefits and perceived risks indices exhibited wide dispersion in country-level means, with differences that were large relative to within-country variability. Three primary profiles emerged. Teachers from countries such as Vietnam, Uzbekistan, Azerbaijan, and Kazakhstan demonstrated optimistic views, perceiving strong benefits and low risks. In contrast, teachers in several English-speaking territories and countries such as Alberta (Canada), Australia, New Zealand, and the United States reported relatively high perceived benefits but also high perceived risks, suggesting awareness of both opportunities and challenges of adopting AI in education. European countries including France and most Nordic countries (Denmark, Finland, Norway, Sweden) exhibited the most cautious profiles characterized by lower benefit perceptions and heightened risk concerns. These patterns indicate that teachers’ perceptions of AI are multidimensional and cannot be reduced to simple pro- or anti-AI stances.

\begin{figure*}[t]
    \centering
    \includegraphics[width=0.88\textwidth]{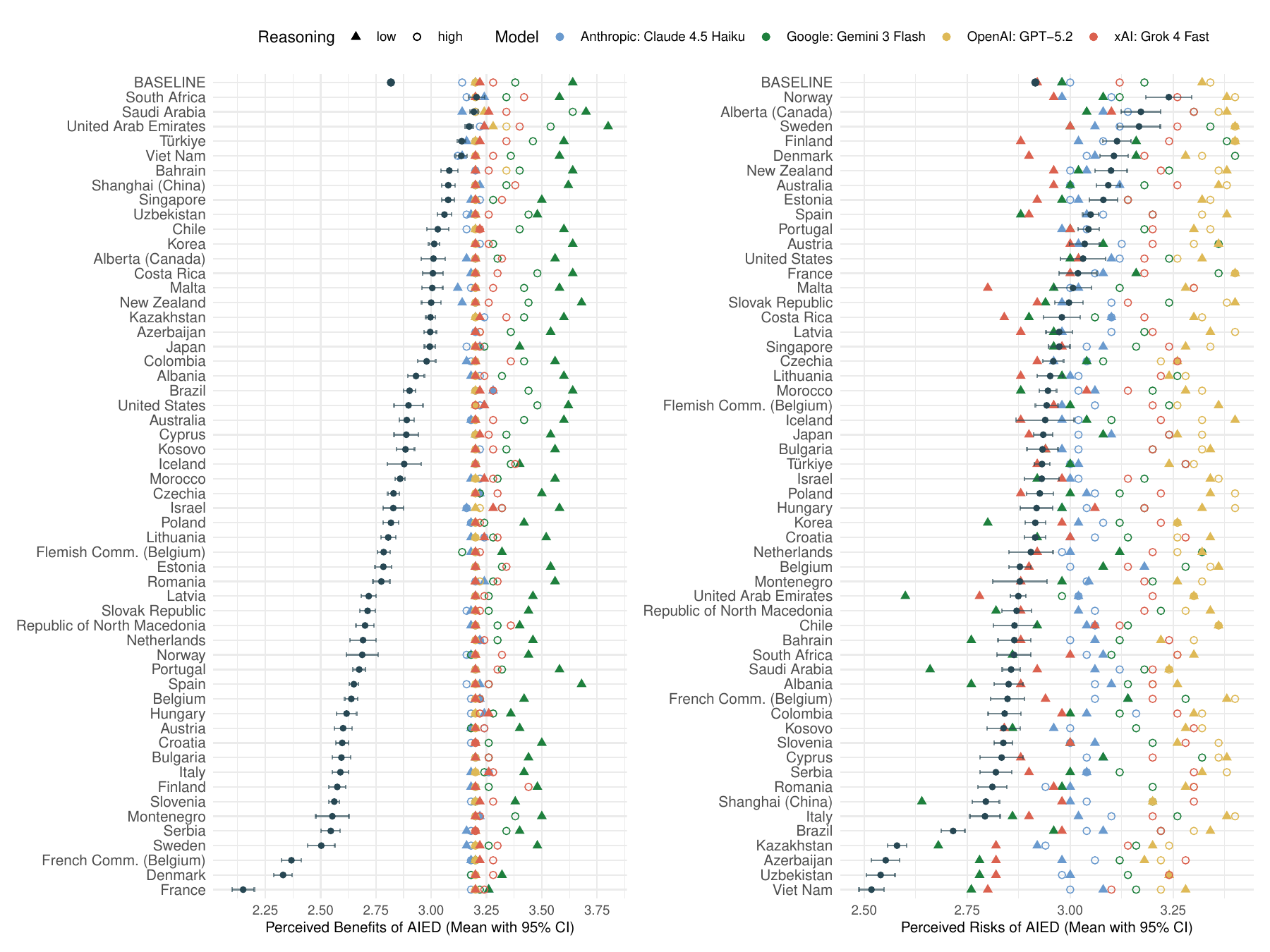}
    \caption{Perceived benefits (left) and risks (right) of AI in education across countries/territories. Black circles show the mean human responses representative for each country with 95\% confidence intervals. Colored shapes show mean LLM responses from four models with triangles for low-reasoning and hollow circles for high-reasoning mode. The BASELINE row shows the overall average of human responses (black circle) and non-country-specific LLM responses (colored shapes).}
    \label{fig:combined_plot}
\end{figure*}

\subsection{Alignment Between LLM Responses and Teacher Perceptions}

Next, we assessed the extent to which LLM-generated responses aligned with teachers’ perceptions of AI benefits and risks at the beginning of the study (i.e., without country-specific prompting as a control strategy). Overall alignment was poor across nearly all models (Figure~\ref{fig:combined_plot}, see baseline row).
For perceived benefits, all four models substantially overstated teachers' perceptions relative to the survey responses. LLM-generated benefit scores approached or exceeded the highest country-level average from the survey, indicating systematic bias toward high perceived AI benefits across all models regardless of reasoning mode. For perceived risks, only Grok 4 Fast none-reasoning accurately represented the overall survey average across countries. Other models skewed toward countries with higher risk perceptions, with GPT-5.2 producing scores that exceeded even the highest country-level average from the survey. LLMs therefore tend to position teachers' AI-related judgments in the upper-right corner of Figure~\ref{fig:2dimension_plot} (high benefits, high risks), which is a profile that no surveyed country actually exhibited.

\begin{figure*}[t]
    \centering
    \includegraphics[width=0.8\textwidth]{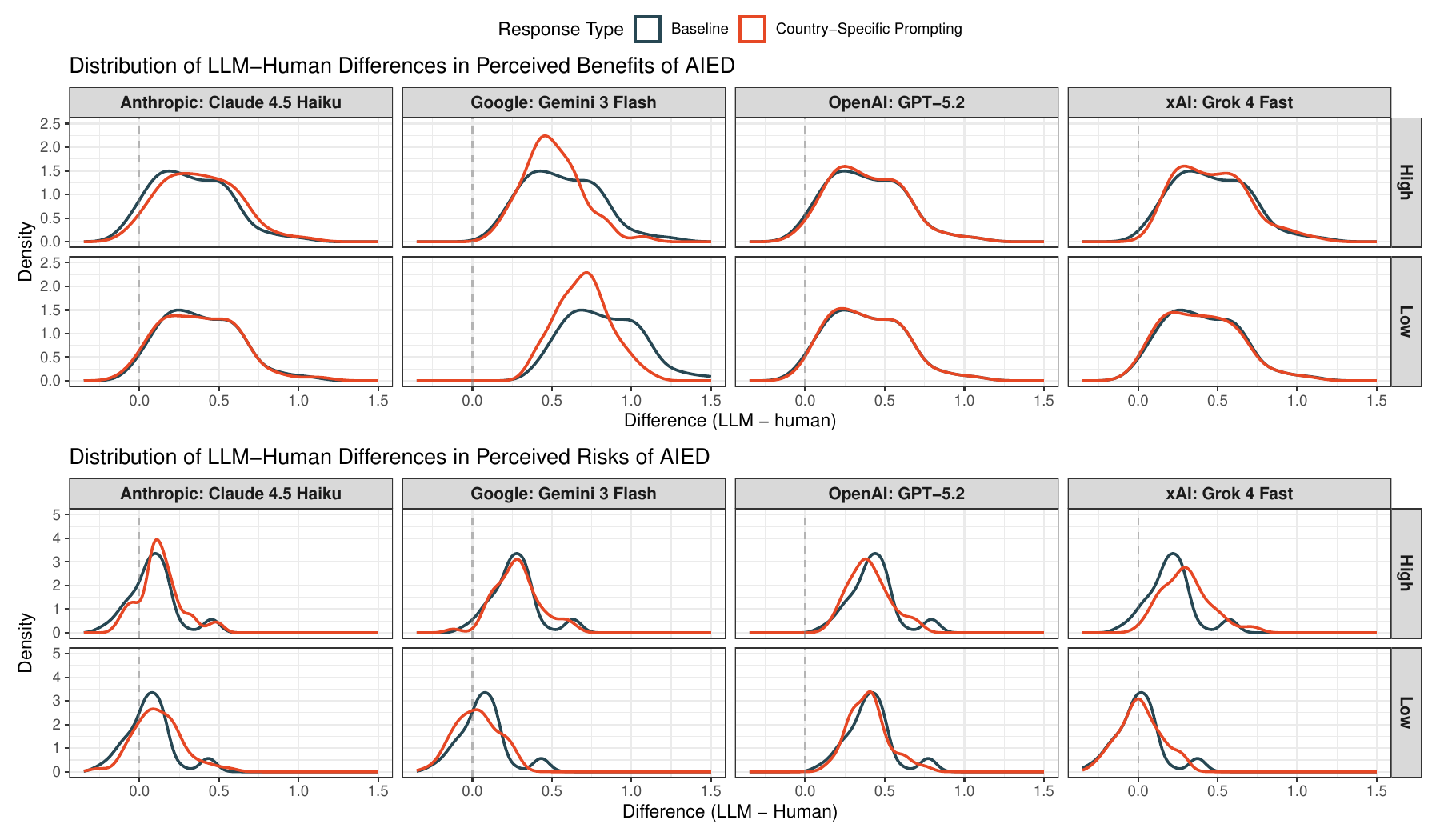}
    \caption{Distribution of differences between LLM and human responses across countries for perceived benefits (top) and risks (bottom) of AI in education. Each panel shows density distributions for four LLM models (columns) with high reasoning mode (top row within each panel) and low/no reasoning mode (bottom row within each panel). Black lines represent differences using baseline LLM responses and red lines with country-specific prompting. A vertical dashed line at zero indicates perfect alignment between LLM and human responses. Positive values indicate that LLMs scored higher than humans.}
    \label{fig:density_differences}
\end{figure*}

\subsection{Effects of Country-Specific Prompting on LLM Alignment}

We evaluated whether country-specific prompting could improve alignment of LLM representations with local teacher perceptions. As show in Figure~\ref{fig:combined_plot} and Figure~\ref{fig:density_differences}, instructing LLMs to respond as teachers from specific countries produced only modest improvement in Gemini 3 Flash (both high and low reasoning) and very small improvement in GPT-5.2 (high reasoning). For other models, country-specific prompting did not substantially reduce deviations from teacher benchmarks, or even introduced additional distortions.

For perceived benefits, all models except Gemini 3 Flash remained anchored to their baseline scores under country-specific prompting, showing minimal cross-national variance. Their country-specific representations therefore diverged substantially from teacher responses for most countries, and demonstrated a persistent bias toward overly positive perception of AI benefits. Among these models, only Grok 4 Fast non-reasoning showed statistically significant reduction in the absolute difference between model country-specific representation and teacher survey estimates (paired Wilcoxon signed-rank test: $V = 856, p < 0.001$), though the magnitude of change was minimal (mean decrease from 0.402 to 0.393). Notably, Claude 4.5 Haiku with high reasoning exhibited a counterproductive effect, with country-specific prompting significantly increasing the absolute LLM-human differences from a mean of 0.328 to 0.379 ($V = 30.5, p < 0.001$). All other non-Gemini model-reasoning combinations showed no statistically significant shifts ($p > 0.05$). 

In contrast, Gemini 3 Flash in both reasoning modes demonstrated greater responsiveness to country-specific prompting, with outputs shifting closer to teacher averages for some Asian countries (e.g., Japan, Singapore) and European countries (e.g., Denmark, France, Norway). Country-specific prompting significantly reduced the absolute LLM-human difference, with mean decreasing from 0.562 to 0.504 for high reasoning ($V = 1192, p < 0.001$) and from 0.822 to 0.696 for low reasoning ($V = 1273, p < 0.001$). However, even these improvements remained insufficient to achieve strong LLM-teacher alignment.

Beyond absolute value shifts, country-specific prompting enabled some LLMs to capture relative ranking patterns in teacher benefit perceptions. Gemini 3 Flash approximated these ranking patterns well, showing strong Spearman correlations with teacher surveys in both high reasoning ($\rho = 0.69, p < 0.001$) and low reasoning ($\rho = 0.71, p < 0.001$). The responses of GPT-5.2 with low reasoning and Grok 4 Fast with high reasoning under country-specific prompting also showed significant but weak correlations with teacher survey results ($\rho \approx 0.3$; Table~\ref{tab:spearman_correlation}).

\begin{table}[ht]
\centering
\small
\caption{Spearman correlations between country-level survey responses and LLM responses with country-specific prompts.} 
\label{tab:spearman_correlation}
\begin{tabular}{lccc}
  \toprule
\textbf{Model} & \textbf{Reasoning} & \textbf{Cor} & \textbf{$p$-value} \\ 
\midrule
\textbf{Perceived Benefits} & & & \\
Claude 4.5 Haiku & High & 0.10 & 0.463 \\ 
  Claude 4.5 Haiku & Low & -0.13 & 0.354 \\ 
  Gemini 3 Flash & High & \textbf{0.69} & \textbf{<0.001} \\ 
  Gemini 3 Flash & Low & \textbf{0.71} & \textbf{<0.001} \\ 
  GPT-5.2 & High & 0.21 & 0.119 \\ 
  GPT-5.2 & Low & \textbf{0.31} & \textbf{0.020} \\ 
  Grok 4 Fast & High & \textbf{0.30} & \textbf{0.027} \\ 
  Grok 4 Fast & Low & 0.04 & 0.774 \\   
  \midrule
\textbf{Perceived Risks} & & & \\
Claude 4.5 Haiku & High & \textbf{0.29} & \textbf{0.033} \\ 
  Claude 4.5 Haiku & Low & 0.03 & 0.829 \\ 
  Gemini 3 Flash & High & 0.26 & 0.060 \\ 
  Gemini 3 Flash & Low & \textbf{0.51} & \textbf{<0.001} \\ 
  GPT-5.2 & High & \textbf{0.41} & \textbf{0.002} \\ 
  GPT-5.2 & Low &\textbf{0.45 }& \textbf{<0.001} \\ 
  Grok 4 Fast & High & -0.07 & 0.618 \\ 
  Grok 4 Fast & Low & \textbf{0.29} & \textbf{0.033} \\ 
   \bottomrule
\end{tabular}
\end{table}
For perceived risks, country-specific prompting produced greater response variance than for benefits, but most models' country-specific representations clustered around their baseline scores within a range much smaller than the range observed in teacher survey responses. Only GPT-5.2 with high reasoning showed significant reductions in absolute LLM-human differences ($V = 912, p = 0.019$), with a mean decrease from 0.425 to 0.402. Conversely, country-specific prompting significantly increased LLM-human differences for Claude 4.5 Haiku in both high reasoning (mean from 0.130 to 0.152; $V = 267, p = 0.005$) and low reasoning (mean from 0.121 to 0.149; $V = 273, p = 0.003$), as well as for Grok 4 Fast with high reasoning (mean from 0.212 to 0.301; $V = 34, p < 0.001$). Other model-reasoning combinations showed no significant shifts. 

Despite accurately representing the overall human average in risk perceptions under baseline prompting, Grok 4 Fast non-reasoning showed the smallest deviations from survey responses (Figure~\ref{fig:density_differences}) but only weakly captured country-level ranking patterns (Spearman $\rho = 0.29, p = 0.033$).
Gemini 3 Flash with low reasoning also closely approximated the overall human average in the baseline condition, with further (though non-significant) reductions in deviations under country-specific prompting and moderate capture of the cross-country rank variation ($\rho = 0.51, p < 0.001$). 
Claude 4.5 Haiku with high reasoning and GPT-5.2 (with both high and low reasoning) demonstrated weak to moderate capture of the country-level ranking patterns (Table~\ref{tab:spearman_correlation}).

Overall, these findings indicate limited steerability in this domain: While prompting models with national teacher identities enabled some models to capture relative differences across countries in teachers' perceived benefits and risks of AI in education, this approach was insufficient to reliably align LLM responses with empirically observed teacher perceptions. Most models continued to systematically overestimate both perceived benefits and risks across most countries.


\subsection{Does Explicit Reasoning Improve Alignment or Steerability?}

We compared reasoning and non-reasoning models within providers to examine whether explicit reasoning capabilities improved alignment or steerability. Contrary to expectations, reasoning models did not consistently outperform non-reasoning models in aligning with teacher survey data. Most models exhibited similar or greater deviations from teacher responses in high reasoning mode, particularly with respect to perceived risks (Figure~\ref{fig:combined_plot}, top row).

Moreover, reasoning capabilities did not systematically enhance responsiveness to country-specific prompting. While reasoning models often produced more elaborated justifications in free-form text, these did not translate into improved alignment and sometimes even enlarge the misalignment with teachers’ benefit–risk judgments. Table~\ref{tab:t_test} shows the results of paired t-test on the country-level LLM-human differences (absolute values) between high and low reasoning for each model. For perceived benefits, only Gemini 3 Flash demonstrated significantly reduced deviations from teacher benchmarks in high reasoning mode ($t = -17.95, p < 0.001$) while Grok 4 Fast demonstrated significantly increased LLM-human difference with high reasoning ($t = 9.18, p < 0.001$). For perceived risks, both Gemini 3 Flash and Grok 4 Fast demonstrated significantly increased deviations from teacher benchmarks. These results suggest that explicit reasoning mechanisms do not, by themselves, resolve alignment challenges when models are asked to represent domain-specific professional perspectives.

\begin{table*}[h]
\centering
\small
\caption{Paired t-tests of human-LLM difference scores (absolute values) comparing high vs. low reasoning, both with country-specific prompting. All comparisons confirmed by Wilcoxon signed-rank tests with consistent significance levels.}
\label{tab:t_test}
\begin{tabular}{lccccccc}
\toprule
 & \multicolumn{2}{c}{\textbf{High Reasoning}} & \multicolumn{2}{c}{\textbf{Low Reasoning}} & & & \\
\cmidrule(lr){2-3} \cmidrule(lr){4-5}
\textbf{Model} & \textbf{Mean} & \textbf{(SD)} & \textbf{Mean} & \textbf{(SD)} & \textbf{$t$ statistic} & \textbf{df} & \textbf{$p$-value} \\ 
\midrule
\textbf{Perceived Benefits} &  &  &  &  &  & & \\
  Claude 4.5 Haiku & 0.379 & (0.230) & 0.378 & (0.237) & 0.21 & 54 & 0.832 \\ 
  Gemini 3 Flash Preview & \textbf{0.504} & (0.182) & \textbf{0.696} & (0.167) & -17.95 & 54 & \textbf{<0.001} \\ 
  GPT 5.2 & 0.388 & (0.226) & 0.384 & (0.229) & 1.38 & 54 & 0.175 \\ 
  Grok 4- Fast & \textbf{0.465} & (0.221) & \textbf{0.393} & (0.232) & 9.18 & 54 & \textbf{<0.001} \\  
\midrule
\textbf{Perceived Risks} &  &  &  & \\
  Claude 4.5 Haiku & 0.152 & (0.113) & 0.149 & (0.122) & 0.34 & 54 & 0.734 \\ 
  Gemini 3 Flash Preview & \textbf{0.287} & (0.138) & \textbf{0.111} & (0.078) & 10.53 & 54 & \textbf{<0.001} \\ 
  GPT 5.2 & 0.402 & (0.129) & 0.398 & (0.127) & 0.69 & 54 & 0.490 \\ 
  Grok 4 Fast & \textbf{0.301} & (0.153) & \textbf{0.105} & (0.084) & 9.47 & 54 & \textbf{<0.001} \\ 
\bottomrule
\end{tabular}
\end{table*}

\section{Discussion}
This study examined cross-national variation in teachers’ perceived benefits and risks of AI in education and evaluated whether LLMs align with these empirically observed perceptions. By combining representative international survey data with systematic model benchmarking, the findings contribute to research on teachers’ trust in educational AI, cross-cultural technology adoption, and AI alignment science. Overall, three main results emerge. First, teachers’ perceived benefits and risks of AI vary substantially across countries (RQ1). Second, most LLMs overstate levels of perceived benefits and risks under baseline prompt that does not specify a national identity (RQ2). Third, neither country-specific prompting nor high reasoning consistently improve LLM alignment, indicating a lack of model steerability (RQ3–RQ4). These findings have implications for teacher use, education research, AI policy, and the growing use of generative AI models as proxies for human views.

\subsection{Cross-National Differences in Teachers’ AI Benefit-Risk Judgments}

Our findings across 55 countries and territories lend empirical support to prior work conceptualizing teachers’ trust in AI-enabled educational technologies as grounded in assessments of anticipated benefits and perceived risks \cite{viberg2024trust, nazaretsky2022teachers}. Consistent with this literature, teachers did not exhibit uniformly positive or negative attitudes toward AI. Instead, perceived benefits and risks coexisted, suggesting that teachers evaluate AI pragmatically in relation to instructional utility, professional autonomy, student learning, and ethical considerations. This supports perspectives framing trust in AI not as general acceptance or resistance but as a negotiated balance between opportunities and concerns \cite{afroogh2024trust}.

At the same time, the magnitude of cross-national variation observed in the presents work extends previous studies that relied on single-country or small multi-country samples \cite{filiz2025teachers, kim2025perceptions, calleja2025primary}. This variance was evident across the full sample, not confined to a few countries, highlighting the importance of grounding global discussions of educational AI in representative, cross-national data rather than extrapolating from single-country studies. The results confirm calls for large-scale comparative evidence in AIED, EDM, and Learning at Scale research \cite{viberg2024trust,baker2019culture}, showing that teachers’ perceptions are embedded in national educational systems, professional norms, and policy environments. Such contextual variation is consistent with broader research on cultural differences in technology adoption and trust in automation \cite{leidner2006review, chien2018culture}. Importantly, the observed heterogeneity suggests that global narratives about AI in education risk oversimplifying teacher perspectives if they rely on limited or geographically narrow evidence.

\subsection{Limited Alignment between LLM Outputs and Teacher Perceptions}

Despite increasing use of LLMs to simulate or approximate stakeholder perspectives \cite{jiang2025simulating, matz2024potential, yao2024value}, our findings indicate limited alignment between models' outputs and empirically observed teacher perceptions. Across models, responses tended to overestimate both perceived benefits and risks. This pattern extends prior alignment research demonstrating systematic cultural biases in LLM outputs \cite{tao2024cultural, khan2025randomness, atari2023which}, moving beyond abstract cultural values to domain-specific professional judgments that are directly relevant to educational policy and practice. This finding highlights a fundamental difference in how LLMs and human stakeholders form perspectives: teachers' responses reflect professional knowledge, personal experience, and classroom realities, whereas LLM outputs are shaped by patterns in public discourse (which may amplify optimism or concern about AI in education) and by human feedback most likely from non-professionals incorporated during model training. Caution is therefore warranted when using LLMs to simulate stakeholder perspectives in domain-specific professional contexts such as AI adoption in education. Echoing prior work in AI Ethics and Society and AI for Culture (e.g., \cite{seth2025deep, prabhakaran2022culturalincongruenciesartificialintelligence}), our findings highlights the need to diversify training data and to advance the mathematical and algorithmic foundations of alignment. We further argue that mitigating representational bias requires moving beyond sociodemographic diversity to recognize professional-domain representation as an equally critical dimension of alignment.


\subsection{Limits of Prompt-Based and Reasoning Steerability}
Prior work in AI alignment suggests that prompt-based strategies, including cultural or identity prompting, can partially improve alignment between model outputs and human values \cite{tao2024cultural}. In contrast, our results indicate that country-specific prompting produced only limited changes in LLM responses and did not reliably recover empirically observed teacher perceptions. This suggests that domain-specific professional judgments may be less amenable to prompt-based steering than more general value statements. One possible explanation is that teachers’ perceptions of AI are shaped by professional experience, institutional constraints, and education policy contexts that are not easily inferred from simple national identity cues. 
Future research should examine whether richer contextual prompting (e.g., specifying school technological infrastructure, teachers’ AI literacy, or pedagogical philosophy) can improve alignment between LLM representations and human professional perspectives. Such work would advance alignment in professional judgment settings from a user-centered standpoint, complementing ongoing efforts in model-level technical development.

The compression of cross-national variation observed in model outputs under country-specific prompting is particularly consequential. Education policy discussions increasingly rely on international comparisons to tailor AI strategies to local conditions. Our findings show that LLM-generated representations may inadvertently smooth over meaningful differences between educational contexts. Such homogenization could contribute to one-size-fits-all AI policy narratives that do not reflect locally salient concerns among educators. Meanwhile, Spearman correlation analyses indicated that some LLMs with less compression of cross-national variation could moderately capture the relative ranking patterns of cross-national variation in perceived benefits and risks under country-specific prompting. This finding suggests that while LLMs are unreliable for estimating population means or effect sizes, some of them have the potential to serve as supplementary tools for exploratory tasks such as identifying cross-national narratives or generating hypotheses about perceived differences across countries.

Recent LLM development increasingly emphasizes explicit reasoning capabilities as a means of improving reliability, coherence, and alignment. However, our within-provider comparisons showed that reasoning-enabled models did not consistently exhibit higher alignment with teacher perceptions or greater responsiveness to country-specific prompting. In some cases, reasoning models diverged further from teacher benchmarks, particularly regarding perceived risks.
These results suggest that reasoning capabilities may enhance internal information processing without necessarily improving alignment with empirically observed human judgments. When the target construct reflects latent professional attitudes rather than factual knowledge, reasoning processes alone may not compensate for gaps in training data or contextual understanding. This finding contributes to ongoing discussions about the relationship between reasoning performance and alignment in generative AI systems (e.g., \cite{mcintosh2024reasoning,kierans2025renorm}).

\subsection{Implications}
The findings have several implications for the use of AI in large-scale educational transformation. Most fundamentally, the substantial cross-national variation in teachers’ perceived benefits and risks suggests that scaling AI in education cannot rely on uniform assumptions about teacher readiness or acceptance. Teachers’ trust in AI remains context-dependent, shaped by professional norms, institutional environments, and national policy contexts. Efforts to scale AI-enabled learning systems therefore require locally grounded implementation strategies rather than universal deployment models.

Echoing LLM audit in domains such as psychology \cite{almeida2024exploring}, cognitive science \cite{gao2025take}, and demographic identity research \cite{wang2025large}, the limited alignment between LLM outputs and empirically observed teacher perceptions also raises caution about using generative AI as a substitute for stakeholder consultation in large-scale educational initiatives. While LLMs can rapidly synthesize information and simulate opinions, they may obscure meaningful contextual variation or introduce systematic distortions. In learning-at-scale contexts, researchers should remain alert to such misrepresentations to avoid inappropriate technology design, professional development strategies, and policy prioritization. Beyond research and policy implications, this misalignment has practical implications for teachers’ use of AI: teachers should critically evaluate AI suggestions, as they may be context-insensitive, confusing, or even misleading.

At the same time, our findings suggest a potential complementary role for LLMs in global studies of educators' perceptions of AI. Researchers could use generative AI to support scalable analysis, hypothesis generation, or scenario exploration regarding cross-national patterns in perceived benefits and risks, while relying on representative empirical data to ground decisions about educational AI adoption. For example, LLMs may help generate hypotheses about how policies or interventions shape teachers’ adoption of AI across countries, and identify “typical” or contrasting educational system profiles to prioritize cases for comparative analysis or survey design. In this way, LLMs can function as exploratory tools that inform research planning, while empirical data from teachers remain indispensable for valid inference.

Finally, the finding that explicit reasoning capabilities do not necessarily improve alignment highlights an important consideration for scalable AI deployment: advances in model sophistication do not automatically translate into better representation of educator perspectives. Responsible scaling therefore requires ongoing empirical validation of AI tools against authentic stakeholder data, alongside careful algorithm development and evaluation.
\subsection{Limitations and Future Study}

This study contributes to understanding whether LLMs can serve as scalable proxies for teacher perspectives on AI in education. However, several limitations constrain how these findings should be interpreted and informs future directions or research.

\textit{First}, we selected the TALIS dataset as the benchmark because it provides large-scale, cross-national evidence on teachers’ perceived benefits and risks of AI with rigorous sampling strategies and weighting procedures~\cite{oecd2024talis_technical}; nonetheless, the uneven distribution of teachers across educational levels may limit its use for auditing LLM representations of generalized teacher views. Additionally, this study focused on perception items in TALIS rather than enacted instructional practices, leaving open how these attitudes translate into adoption, classroom integration, and learning outcomes. 
Future studies should link perception data with measures of actual AI use and student outcomes to better understand how teachers perceive and respond to AI-driven educational transformation.

\textit{Second}, our operationalization of teacher perceptions focuses on benefit–risk judgments as core antecedents of trust in AI in education. While this framework is theoretically grounded in prior educational AI research~\cite{nazaretsky2022teachers, viberg2024trust}, large-scale adoption also depends on factors such as institutional capacity, professional development access, AI literacy, infrastructure, and policy alignment~\cite{taheri2025factors}. These structural conditions are particularly salient when considering scalable AI deployment across education systems and cultural contexts. Future research should further identify these factors to inform more context-sensitive scaling strategies.

\textit{Third}, the LLM benchmarking reflects model capabilities at a specific technological moment. As generative AI systems rapidly evolve, improvements in data, reasoning, and alignment may change their ability to simulate stakeholder perspectives. Accordingly, the observed misalignment reflects current limitations rather than stable properties of generative AI.

\textit{Fourth}, prompting LLMs to represent national teacher perspectives necessarily simplifies complex educational contexts. Teachers’ attitudes toward AI are shaped by curriculum mandates, governance structures, assessment regimes, and cultural expectations about teaching and technology~\cite{viberg2024trust}. Moreover, TALIS items were administered in local survey languages, whereas our steering used a uniform English identity cue. We chose English prompting because it reflects common practice in cross-national research and allowed output verification across countries. However, this language mismatch may have altered item interpretation and response framing, potentially attenuating country-specific effects and contributing to the observed compression of cross-national differences in model outputs. Further work is needed to refine prompting strategies, develop robust benchmarking frameworks, and pursue participatory validation with educators.


Overall, this study highlights that scaling AI in education is not merely a technical challenge, but an inherently socio-educational one. Substantial cross-national variation in teachers’ perceptions of AI benefits and risks underscores the need for locally grounded understanding of educators’ perspectives and context-sensitive design of AI technologies and policies. At the same time, the limited reflection of these differences in current LLM outputs suggests that relying on generative AI alone may flatten important contextual variation. Teachers’ perspectives remain central to trust, adoption, and meaningful classroom integration. While generative AI can support comparative analysis and hypothesis generation at scale, direct engagement with educators remains essential for responsible, context-sensitive AI-in-education policy and practice.
\begin{acks}
This material is based upon work supported by the National Science Foundation under Grant No. 2237593. Viberg's work is supported by grant 2025-04552 from the Swedish Research Council.
\end{acks}

\bibliographystyle{ACM-Reference-Format}
\bibliography{ref}

@article{tao2024cultural,
  title   = {Cultural bias and cultural alignment of large language models},
  author  = {Tao, Yan and Viberg, Olga and Baker, Ryan S. and Kizilcec, Ren{\'e} F.},
  journal = {PNAS Nexus},
  year    = {2024},
  volume  = {3},
  number  = {9},
  pages   = {pgae346},
  doi     = {10.1093/pnasnexus/pgae346}
}

@article{viberg2024trust,
  title   = {What Explains Teachers' Trust of AI in Education across Six Countries?},
  author  = {Viberg, Olga and Cukurova, Mutlu and Feldman-Maggor, Yael and Alexandron, Giora and Shirai, Shizuka and Kanemune, Susumu and Wasson, Barbara and T{\o}mte, Cathrine and Spikol, Daniel and Milrad, Marcelo and Coelho, Raquel and Kizilcec, Ren{\'e} F.},
  journal={International Journal of Artificial Intelligence in Education},
  volume={35},
  number={3},
  pages={1288--1316},
  year={2025},
  publisher={Springer}
}

@article{garcia2025perceptions,
  title={Perceptions, strategies, and challenges of teachers in the integration of artificial intelligence in primary education: A systematic review},
  author={Garcia, Olga Arranz and Garc{\'\i}a, Mar{\'\i}a del Carmen Romero and Alonso-Secades, Vidal},
  journal={Journal of Information Technology Education: Research},
  volume={24},
  pages={006},
  year={2025},
  publisher={Informing Science Institute. 131 Brookhill Court, Santa Rosa, CA 95409}
}

@article{shen2025psychological,
  title={Psychological safety and trust as drivers of teachers’ continued use of AI tools in classrooms},
  author={Shen, Lijuan and Qiu, Nengsheng and Wang, Zhanfeng},
  journal={Scientific Reports},
  volume={15},
  number={1},
  pages={31426},
  year={2025},
  publisher={Nature Publishing Group UK London}
}

@misc{delikoura2025superficial,
  title={From superficial outputs to superficial learning: Risks of large language models in education},
  author={Delikoura, Iris and Fung, Yi R and Hui, Pan},
  year={2025},
  eprint={2509.21972},
  archivePrefix={arXiv},
  primaryClass={cs.CY},
  url={https://doi.org/10.48550/arXiv.2509.21972
}
}

@article{huang2025artificial,
  title={Artificial intelligence in K-12 education: An umbrella review},
  author={Huang, Ruiping and Yin, Yue and Zhou, Na and Lang, Fei},
  journal={Computers and Education: Artificial Intelligence},
  volumn = {10},
  pages={100519},
  year={2025},
  publisher={Elsevier}
}

@article{kizilcec2024advance,
  title={To advance AI use in education, focus on understanding educators},
  author={Kizilcec, Ren{\'e} F},
  journal={International Journal of Artificial Intelligence in Education},
  volume={34},
  number={1},
  pages={12--19},
  year={2024},
  publisher={Springer}
}

@article{tan2025artificial,
  title={Artificial intelligence in teaching and teacher professional development: A systematic review},
  author={Tan, Xiao and Cheng, Gary and Ling, Man Ho},
  journal={Computers and Education: Artificial Intelligence},
  volume={8},
  pages={100355},
  year={2025},
  publisher={Elsevier}
}

@article{nazaretsky2022teachers,
  title={Teachers' trust in AI-powered educational technology and a professional development program to improve it},
  author={Nazaretsky, Tanya and Ariely, Moriah and Cukurova, Mutlu and Alexandron, Giora},
  journal={British journal of educational technology},
  volume={53},
  number={4},
  pages={914--931},
  year={2022},
  publisher={Wiley Online Library}
}

@article{kim2025perceptions,
  title={Perceptions and preparedness of K-12 educators in adopting generative AI},
  author={Kim, Juhee},
  journal={Research in Learning Technology},
  volume = {33},
  pages = {3448},
  year = {2025},
  doi = {10.25304/rlt.v33.3448},
  url = {http://dx.doi.org/10.25304/rlt.v33.3448}
}

@article{calleja2025primary,
  author = {Calleja, James and Camilleri, Patrick},
  title = {Primary school teachers' perceptions towards the use of generative AI in teaching using lesson study},
  journal = {International Journal for Lesson and Learning Studies},
  year = {2025},
  volume = {14},
  number = {3},
  pages = {237--252},
  doi = {10.1108/IJLLS-11-2024-0268},
  url = {https://doi.org/10.1108/IJLLS-11-2024-0268}
}

@techreport{oecd2025talis,
  author = {{OECD}},
  title = {Teaching and Learning International Survey (TALIS) 2024 Conceptual Framework},
  institution = {OECD Publishing},
  address = {Paris},
  year = {2025},
  doi = {10.1787/7b8f85d4-en},
  url = {https://doi.org/10.1787/7b8f85d4-en}
}

@article{baker2019culture,
  title={Culture in Computer-Based Learning Systems: Challenges and Opportunities.},
  author={Baker, Ryan S and Ogan, Amy E and Madaio, Michael and Walker, Erin},
  journal={Computer-Based Learning in Context},
  volume={1},
  number={1},
  pages={1--13},
  year={2019},
  publisher={ERIC}
}

@inproceedings{yao2024value,
  title = "Value {FULCRA}: Mapping Large Language Models to the Multidimensional Spectrum of Basic Human Value",
    author = "Yao, Jing  and
      Yi, Xiaoyuan  and
      Gong, Yifan  and
      Wang, Xiting  and
      Xie, Xing",
    editor = "Duh, Kevin  and
      Gomez, Helena  and
      Bethard, Steven",
    booktitle = "Proceedings of the 2024 Conference of the North American Chapter of the Association for Computational Linguistics: Human Language Technologies (Volume 1: Long Papers)",
    month = jun,
    year = "2024",
    address = "Mexico City, Mexico",
    publisher = "Association for Computational Linguistics",
    url = "https://aclanthology.org/2024.naacl-long.486/",
    doi = "10.18653/v1/2024.naacl-long.486",
    pages = "8762--8785"
}

@article{gouseti2025ethics,
  title={The ethics of using AI in K-12 education: A systematic literature review},
  author={Gouseti, Anastasia and James, Fiona and Fallin, Lee and Burden, Kevin},
  journal={Technology, Pedagogy and Education},
  volume={34},
  number={2},
  pages={161--182},
  year={2025},
  publisher={Taylor \& Francis}
}

@article{filiz2025teachers,
  title={Teachers and AI: Understanding the factors influencing AI integration in K-12 education},
  author={Filiz, Ozan and Kaya, Mehmet Haldun and Adiguzel, Tufan},
  journal={Education and Information Technologies},
  volume = {30},
  pages = {17931--17967},
  year = {2025},
  doi = {10.1007/s10639-025-13463-2},
  url = {https://doi.org/10.1007/s10639-025-13463-2}
}

@inproceedings{khan2025randomness,
  author = {Khan, Ariba and Casper, Stephen and Hadfield-Menell, Dylan},
title = {Randomness, Not Representation: The Unreliability of Evaluating Cultural Alignment in LLMs},
year = {2025},
isbn = {9798400714825},
publisher = {Association for Computing Machinery},
address = {New York, NY, USA},
url = {https://doi.org/10.1145/3715275.3732147},
doi = {10.1145/3715275.3732147},
booktitle = {Proceedings of the 2025 ACM Conference on Fairness, Accountability, and Transparency},
pages = {2151–2165},
numpages = {15},
location = {
},
series = {FAccT '25}
}

@article{taheri2025factors,
  title={Factors Influencing Educators' AI Adoption: A Grounded Meta-Analysis Review},
  author={Taheri, Rana and Nazemi, Neda and Pennington, Sarah E and Clark, Jason A and Dadgostari, Faraz},
  journal={Computers and Education: Artificial Intelligence},
  pages={100464},
  year={2025},
  doi = {10.1080/23812346.2025.2603822},
  url = {https://doi.org/10.1080/23812346.2025.2603822},
  publisher={Elsevier}
}

@article{jiang2025simulating,
  title={Simulating social perception with large language models: perceptions of China’s common prosperity},
  author={Jiang, Zhuoren and Huang, Biao and Ge, Jianan and Lin, Chenxi and Xu, Yueqian and Yu, Jianxing},
  journal={Journal of Chinese Governance},
  pages={1--29},
  year={2025},
  publisher={Taylor \& Francis}
}

@article{matz2024potential,
  title={The potential of generative AI for personalized persuasion at scale},
  author={Matz, Sandra C and Teeny, Jacob D and Vaid, Sumer S and Peters, Heinrich and Harari, Gabriella M and Cerf, Moran},
  journal={Scientific Reports},
  volume={14},
  number={1},
  pages={4692},
  year={2024},
  publisher={Nature Publishing Group UK London}
  
}

@article{karran2025multi,
  title={Multi-stakeholder perspective on responsible artificial intelligence and acceptability in education},
  author={Karran, Alexander John and Charland, Patrick and Trempe-Martineau, Jo{\'e} and Ortiz de Guinea Lopez de Arana, Ana and Lesage, Anne-Marie and Senecal, Sylvain and Leger, Pierre-Majorique},
  journal={npj Science of Learning},
  volume={10},
  number={1},
  pages={44},
  year={2025},
  publisher={Nature Publishing Group UK London}
}

@article{afroogh2024trust,
  title={Trust in AI: progress, challenges, and future directions},
  author={Afroogh, Saleh and Akbari, Ali and Malone, Emmie and Kargar, Mohammadali and Alambeigi, Hananeh},
  journal={Humanities and Social Sciences Communications},
  volume={11},
  number={1},
  pages={1--30},
  year={2024},
  publisher={Palgrave}
}

@article{mcintosh2024reasoning,
  title={A reasoning and value alignment test to assess advanced gpt reasoning},
  author={McIntosh, Timothy R and Liu, Tong and Susnjak, Teo and Watters, Paul and Halgamuge, Malka N},
  journal={ACM Transactions on Interactive Intelligent Systems},
  volume={14},
  number={3},
  pages={1--37},
  year={2024},
  publisher={ACM New York, NY}
}

@misc{kierans2025renorm,
  author = {Kierans, Aidan and Dori-Hacohen, Shiri},
  title = {Position: Aligning {AI} requires automating reasoning norms},
  year = {2026},
  howpublished = {SSRN},
  note = {Available at SSRN},
  doi = {10.2139/ssrn.5399451},
  url = {https://ssrn.com/abstract=5399451}
}

@book{holmes2023guidance,
  title={Guidance for generative AI in education and research},
  author={Holmes, Wayne and Miao, Fengchun and others},
  year={2023},
  publisher={Unesco Publishing}
}

@article{celik2022promises,
  title   = {The promises and challenges of artificial intelligence for teachers: A systematic review of research},
  author  = {Celik, Ismail and Dindar, Muhterem and Muukkonen, Hanni and J{\"a}rvel{\"a}, Sanna},
  journal = {TechTrends},
  year    = {2022},
  volume  = {66},
  number  = {4},
  pages   = {616--630},
  doi     = {10.1007/s11528-022-00715-y}
}

@article{celik2025co,
  author = {Celik, Ismail and Kontkanen, Sini and Laru, Jari and Dalyanci, Alanur Ahsen},
  title = {Co-constructing adaptive lesson plans with {GenAI}: Pre-service teachers' {Intelligent-TPACK} and prompt engineering strategies},
  journal = {Computers \& Education},
  volume = {241},
  pages = {105485},
  year = {2026},
  issn = {0360-1315},
  doi = {10.1016/j.compedu.2025.105485},
  url = {https://www.sciencedirect.com/science/article/pii/S0360131525002532}
}

@article{zawacki2019systematic,
  title   = {Systematic review of research on artificial intelligence applications in higher education---where are the educators?},
  author  = {Zawacki-Richter, Olaf and Mar{\'\i}n, Victoria I. and Bond, Melissa and Gouverneur, Franziska},
  journal = {International Journal of Educational Technology in Higher Education},
  year    = {2019},
  volume  = {16},
  number  = {1},
  pages   = {1--27},
  doi     = {10.1186/s41239-019-0171-0}
}

@article{velander2023swedish,
  title   = {Artificial intelligence in K-12 education: eliciting and reflecting on Swedish teachers' understanding of AI and its implications for teaching \& learning},
  author  = {Velander, Johanna and Taiye, M. A. and Otero, N. and Milrad, Marcelo},
  journal = {Education and Information Technologies},
  volume = {29},
  pages = {4085--4105},
  year = {2024},
  doi = {10.1007/s10639-023-11990-4},
  url = {https://doi.org/10.1007/s10639-023-11990-4}
}

@inproceedings{barocas2021disaggregated,
author = {Barocas, Solon and Guo, Anhong and Kamar, Ece and Krones, Jacquelyn and Morris, Meredith Ringel and Vaughan, Jennifer Wortman and Wadsworth, W. Duncan and Wallach, Hanna},
title = {Designing Disaggregated Evaluations of AI Systems: Choices, Considerations, and Tradeoffs},
year = {2021},
isbn = {9781450384735},
publisher = {Association for Computing Machinery},
address = {New York, NY, USA},
url = {https://doi.org/10.1145/3461702.3462610},
doi = {10.1145/3461702.3462610},
booktitle = {Proceedings of the 2021 AAAI/ACM Conference on AI, Ethics, and Society},
pages = {368–378},
numpages = {11},
location = {Virtual Event, USA},
series = {AIES '21}
}

@unpublished{sandvig2014auditing,
  title     = {Auditing algorithms: research methods for detecting discrimination on internet platforms},
  author    = {Sandvig, Christian and Hamilton, Kevin and Karahalios, Karrie and Langbort, Cedric},
  year = {2014},
  month = May,
  note = {Paper presented at "Data and Discrimination: Converting Critical Concerns into Productive Inquiry," preconference at the 64th Annual Meeting of the International Communication Association, Seattle, WA, USA}
}

@inproceedings{arora2023probing,
    title = "Probing Pre-Trained Language Models for Cross-Cultural Differences in Values",
    author = "Arora, Arnav  and
      Kaffee, Lucie-aim{\'e}e  and
      Augenstein, Isabelle",
    editor = "Dev, Sunipa  and
      Prabhakaran, Vinodkumar  and
      Adelani, David Ifeoluwa  and
      Hovy, Dirk  and
      Benotti, Luciana",
    booktitle = "Proceedings of the First Workshop on Cross-Cultural Considerations in NLP (C3NLP)",
    month = may,
    year = "2023",
    address = "Dubrovnik, Croatia",
    publisher = "Association for Computational Linguistics",
    url = "https://aclanthology.org/2023.c3nlp-1.12/",
    doi = "10.18653/v1/2023.c3nlp-1.12",
    pages = "114--130"
}

@inproceedings{naous2023beer,
  title     = {Having beer after prayer? Measuring cultural bias in large language models},
  author    = {Naous, Tarek and Ryan, Michael J. and Ritter, Alan and Xu, Wei},
  booktitle = {Proceedings of the 62nd Annual Meeting of the Association for Computational Linguistics (Volume 1: Long Papers)},
  year      = {2024},
  pages     = {16366--16393},
  publisher = {Association for Computational Linguistics},
  address = {Bangkok, Thailand}
}

@article{holmes2022state,
  title={State of the art and practice in AI in education},
  author={Holmes, Wayne and Tuomi, Ilkka},
  journal={European journal of education},
  volume={57},
  number={4},
  pages={542--570},
  year={2022},
  publisher={Wiley Online Library}
}

@article{leidner2006review,
  author  = {Leidner, Dorothy E. and Kayworth, Timothy},
  title = {A review of culture in information systems research: Toward a theory of information technology culture conflict},
  journal = {MIS Quarterly},
  volume = {30},
  number = {2},
  pages = {357--399},
  year = {2006},
  doi = {10.2307/25148735},
  url = {https://doi.org/10.2307/25148735}
}

@article{chien2018culture,
  title   = {The effect of culture on trust in automation: reliability and workload},
  author  = {Chien, Shu-Yang and Lewis, Michael and Sycara, Katia and Liu, Jui-Shiang and Kumru, Asli},
  journal = {ACM Transactions on Interactive Intelligent Systems},
  year    = {2018},
  volume  = {8},
  number  = {4},
  pages   = {1--31},
  doi     = {10.1145/3230736}
}

@techreport{oecd2024talis_technical,
  author       = {{OECD}},
  booktitle    = {Results from {TALIS} 2024: The State of Teaching},
  institution  = {OECD Publishing},
  year         = {2024},
  address      = {Paris},
  url          = {https://www.oecd.org/en/publications/results-from-talis-2024_90df6235-en/full-report}
}

@inproceedings{tripathi2025teaching,
  title={Teaching and learning with AI: a qualitative study on K-12 teachers' use and engagement with artificial intelligence},
  author={Tripathi, Tarang and Sharma, Smriti R and Singh, Vatsala and Bhargava, Palaash and Raj, Chandraditya},
  booktitle={Frontiers in Education},
  volume={10},
  pages={1651217},
  year={2025},
  organization={Frontiers}
}

@inproceedings{yin2025responsible,
  author = {Yin, Yaxuan and Karumbaiah, Shamya and Acquaye, Shona},
title = {Responsible AI in Education: Understanding Teachers’ Priorities and Contextual Challenges},
year = {2025},
isbn = {9798400714825},
publisher = {Association for Computing Machinery},
address = {New York, NY, USA},
url = {https://doi.org/10.1145/3715275.3732176},
doi = {10.1145/3715275.3732176},
booktitle = {Proceedings of the 2025 ACM Conference on Fairness, Accountability, and Transparency},
pages = {2705–2727},
numpages = {23},
location = {
},
series = {FAccT '25}
}

@inproceedings{wang-demszky-2023-chatgpt,
    title = "Is {C}hat{GPT} a Good Teacher Coach? Measuring Zero-Shot Performance For Scoring and Providing Actionable Insights on Classroom Instruction",
    author = "Wang, Rose  and
      Demszky, Dorottya",
    editor = {Kochmar, Ekaterina  and
      Burstein, Jill  and
      Horbach, Andrea  and
      Laarmann-Quante, Ronja  and
      Madnani, Nitin  and
      Tack, Ana{\"i}s  and
      Yaneva, Victoria  and
      Yuan, Zheng  and
      Zesch, Torsten},
    booktitle = "Proceedings of the 18th Workshop on Innovative Use of NLP for Building Educational Applications (BEA 2023)",
    month = jul,
    year = "2023",
    address = "Toronto, Canada",
    publisher = "Association for Computational Linguistics",
    url = "https://aclanthology.org/2023.bea-1.53/",
    doi = "10.18653/v1/2023.bea-1.53",
    pages = "626--667"
}

@inproceedings{10.1145/3654777.3676390,
author = {Fan, Haoxiang and Chen, Guanzheng and Wang, Xingbo and Peng, Zhenhui},
title = {LessonPlanner: Assisting Novice Teachers to Prepare Pedagogy-Driven Lesson Plans with Large Language Models},
year = {2024},
isbn = {9798400706288},
publisher = {Association for Computing Machinery},
address = {New York, NY, USA},
url = {https://doi.org/10.1145/3654777.3676390},
doi = {10.1145/3654777.3676390},
booktitle = {Proceedings of the 37th Annual ACM Symposium on User Interface Software and Technology},
articleno = {146},
numpages = {20},
location = {Pittsburgh, PA, USA},
series = {UIST '24}
}

@inproceedings{lu2025understanding,
author = {Lu, Zhuoran and Lim, Gionnieve and Yin, Ming},
title = {Understanding the Effects of Large Language Model (LLM)-driven Adversarial Social Influences in Online Information Spread},
year = {2025},
isbn = {9798400713958},
publisher = {Association for Computing Machinery},
address = {New York, NY, USA},
url = {https://doi.org/10.1145/3706599.3720019},
doi = {10.1145/3706599.3720019},
booktitle = {Proceedings of the Extended Abstracts of the CHI Conference on Human Factors in Computing Systems},
articleno = {555},
numpages = {7},
location = {
},
series = {CHI EA '25}
}

@misc{atari2023which,
  author       = {Atari, Mohammad and Xue, Mona J and Park, Peter S and Blasi, Damián E and Henrich, Joseph},
  title        = {Which Humans?},
  year         = {2023},
  month        = {Sep 22},
  howpublished = {PsyArXiv},
  doi          = {10.31234/osf.io/5b26t},
  url          = {https://doi.org/10.31234/osf.io/5b26t}
}

@misc{fu2023chainofthoughthubcontinuouseffort,
      title={Chain-of-Thought Hub: A Continuous Effort to Measure Large Language Models' Reasoning Performance}, 
      author={Yao Fu and Litu Ou and Mingyu Chen and Yuhao Wan and Hao Peng and Tushar Khot},
      year={2023},
      eprint={2305.17306},
      archivePrefix={arXiv},
      primaryClass={cs.CL},
      url={https://arxiv.org/abs/2305.17306}, 
}

@inproceedings{huang-chang-2023-towards,
    title = "Towards Reasoning in Large Language Models: A Survey",
    author = "Huang, Jie  and
      Chang, Kevin Chen-Chuan",
    editor = "Rogers, Anna  and
      Boyd-Graber, Jordan  and
      Okazaki, Naoaki",
    booktitle = "Findings of the Association for Computational Linguistics: ACL 2023",
    month = jul,
    year = "2023",
    address = "Toronto, Canada",
    publisher = "Association for Computational Linguistics",
    url = "https://aclanthology.org/2023.findings-acl.67/",
    doi = "10.18653/v1/2023.findings-acl.67",
    pages = "1049--1065"
}

@misc{zhou2024selfdiscoverlargelanguagemodels,
      title={Self-Discover: Large Language Models Self-Compose Reasoning Structures}, 
      author={Pei Zhou and Jay Pujara and Xiang Ren and Xinyun Chen and Heng-Tze Cheng and Quoc V. Le and Ed H. Chi and Denny Zhou and Swaroop Mishra and Huaixiu Steven Zheng},
      year={2024},
      eprint={2402.03620},
      archivePrefix={arXiv},
      primaryClass={cs.AI},
      url={https://arxiv.org/abs/2402.03620}, 
}

@misc{openai2024learning,
  author       = {{OpenAI}},
  title        = {Learning to Reason with {LLMs}},
  year         = {2024},
  month        = sep,
  howpublished = {\url{https://openai.com/index/learning-to-reason-with-llms/}},
  note         = {Accessed: 2026-02-12}
}

@article{khamassi2024strong,
  title={Strong and weak alignment of large language models with human values},
  author={Khamassi, Mehdi and Nahon, Marceau and Chatila, Raja},
  journal={Scientific Reports},
  volume={14},
  number={1},
  pages={19399},
  year={2024},
  publisher={Nature Publishing Group UK London}
}

@article{seth2025deep,
  title={How deep is representational bias in llms? the cases of caste and religion},
  author={Seth, Agrima and Choudhury, Monojit and Sitaram, Sunayana and Toyama, Kentaro and Vashistha, Aditya and Bali, Kalika},
  journal={Proceedings of the AAAI/ACM Conference on AI, Ethics, and Society},
  volume={8},
  number={3},
  pages={2319--2330},
  year={2025}
}

@misc{prabhakaran2022culturalincongruenciesartificialintelligence,
      title={Cultural Incongruencies in Artificial Intelligence}, 
      author={Vinodkumar Prabhakaran and Rida Qadri and Ben Hutchinson},
      year={2022},
      eprint={2211.13069},
      archivePrefix={arXiv},
      primaryClass={cs.CY},
      url={https://arxiv.org/abs/2211.13069}, 
}

@article{almeida2024exploring,
  title={Exploring the psychology of LLMs’ moral and legal reasoning},
  author={Almeida, Guilherme FCF and Nunes, Jos{\'e} Luiz and Engelmann, Neele and Wiegmann, Alex and De Ara{\'u}jo, Marcelo},
  journal={Artificial Intelligence},
  volume={333},
  pages={104145},
  year={2024},
  publisher={Elsevier}
}

@inproceedings{10.1145/3531146.3533233,
author = {Hutchinson, Ben and Rostamzadeh, Negar and Greer, Christina and Heller, Katherine and Prabhakaran, Vinodkumar},
title = {Evaluation Gaps in Machine Learning Practice},
year = {2022},
isbn = {9781450393522},
publisher = {Association for Computing Machinery},
address = {New York, NY, USA},
url = {https://doi.org/10.1145/3531146.3533233},
doi = {10.1145/3531146.3533233},
booktitle = {Proceedings of the 2022 ACM Conference on Fairness, Accountability, and Transparency},
pages = {1859–1876},
numpages = {18},
location = {Seoul, Republic of Korea},
series = {FAccT '22}
}

@article{wang2025large,
  title={Large language models that replace human participants can harmfully misportray and flatten identity groups},
  author={Wang, Angelina and Morgenstern, Jamie and Dickerson, John P},
  journal={Nature Machine Intelligence},
  volume = {7},
  pages = {400--411},
  year = {2025},
  doi = {10.1038/s42256-025-00986-z},
  url = {https://doi.org/10.1038/s42256-025-00986-z}
}

@article{gao2025take,
  title={Take caution in using LLMs as human surrogates},
  author={Gao, Yuan and Lee, Dokyun and Burtch, Gordon and Fazelpour, Sina},
  journal={Proceedings of the National Academy of Sciences},
  volume={122},
  number={24},
  pages={e2501660122},
  year={2025},
  publisher={National Academy of Sciences}
}

@inproceedings{Santurkar_23_whoseopion,
author = {Santurkar, Shibani and Durmus, Esin and Ladhak, Faisal and Lee, Cinoo and Liang, Percy and Hashimoto, Tatsunori},
title = {Whose opinions do language models reflect?},
year = {2023},
publisher = {JMLR.org},
booktitle = {Proceedings of the 40th International Conference on Machine Learning},
articleno = {1244},
numpages = {34},
location = {Honolulu, Hawaii, USA},
series = {ICML'23}
}

@inproceedings{cao-etal-2023-assessing,
    title = "Assessing Cross-Cultural Alignment between {C}hat{GPT} and Human Societies: An Empirical Study",
    author = "Cao, Yong  and
      Zhou, Li  and
      Lee, Seolhwa  and
      Cabello, Laura  and
      Chen, Min  and
      Hershcovich, Daniel",
    editor = "Dev, Sunipa  and
      Prabhakaran, Vinodkumar  and
      Adelani, David Ifeoluwa  and
      Hovy, Dirk  and
      Benotti, Luciana",
    booktitle = "Proceedings of the First Workshop on Cross-Cultural Considerations in NLP (C3NLP)",
    month = may,
    year = "2023",
    address = "Dubrovnik, Croatia",
    publisher = "Association for Computational Linguistics",
    url = "https://aclanthology.org/2023.c3nlp-1.7/",
    doi = "10.18653/v1/2023.c3nlp-1.7",
    pages = "53--67"
}

\end{document}